# BREAK: BRONCHI RECONSTRUCTION BY GEODESIC TRANSFORMATION AND SKELETON EMBEDDING

*Weihao Yu*[1,2], *Hao Zheng*[1,2], *Minghui Zhang*[1,2], *Hanxiao Zhang*[2], *Jiayuan Sun*[3], *Jie Yang*[1,2†]

[1]Institute of Image Processing and Pattern Recognition, Shanghai Jiao Tong University, China
[2]Institute of Medical Robotics, Shanghai Jiao Tong University, Shanghai, China
[3]Department of Respiratory and Critical Care Medicine, Department of Respiratory Endoscopy, Shanghai Chest Hospital, Shanghai Engineering Research Center of Respiratory Endoscopy, China.

## ABSTRACT

Airway segmentation is critical for virtual bronchoscopy and computer-aided pulmonary disease analysis. In recent years, convolutional neural networks (CNNs) have been widely used to delineate the bronchial tree. However, the segmentation results of the CNN-based methods usually include many discontinuous branches, which need manual repair in clinical use. A major reason for the breakages is that the appearance of the airway wall can be affected by the lung disease as well as the adjacency of the vessels, while the network tends to overfit these special patterns in the training set. To learn robust features for these areas, we design a multi-stage framework that adopts the geodesic distance transform to capture the intensity changes between airway lumen and wall. Another reason for the breakages is the intra-class imbalance. Since the volume of the peripheral bronchi may be much smaller than the large branches in an input patch, the common segmentation loss is not sensitive to the breakages among the distal branches. Therefore, in this paper, a breakage-sensitive regularization term is designed and can be easily combined with other loss functions. Extensive experiments are conducted on publicly available datasets. Compared with state-of-the-art methods, our framework can detect more branches while maintaining competitive segmentation performance.

## 1. INTRODUCTION

Airway segmentation is a prerequisite for the automatic pathological analysis of lung diseases, which can effectively reduce the workload of clinicians. It is also the basis of virtual bronchoscopy and surgical simulation navigation.

Recently, methods based on convolutional neural networks (CNNs) [1, 2, 3, 4, 5] are increasingly used in image segmentation tasks. For airway delineation, a major challenge is the breakages among the segmental bronchi. Fig. 1 illustrates three main reasons for this problem. Firstly, the volume of the peripheral bronchi is much smaller than the large airways, which is called intra-class imbalance. If the input patch contains both large and small airways, this problem leads to an insignificant decrease in the training loss when the predictions of the distal small airways become discontinuous. Secondly, some bronchi are close to the vessels whose Hounsfield unit (HU) value is higher than the airway wall in CT images, resulting in inhomogeneous intensities among the airway wall. These uncommon features may not be well learned by the model. Thirdly, the branches that pass through the lesion area or include pathological changes may have irregular shapes. The lumen becomes larger while the wall thickness raises or decreases. These features are also hard to learn during the training.

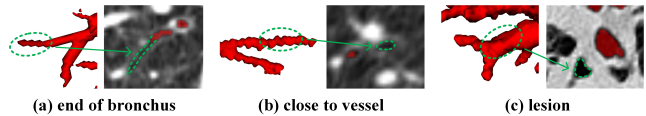

(a) end of bronchus    (b) close to vessel    (c) lesion

**Fig. 1**. Examples of the breakages.

For the first challenge, Wang et al. [6] develop a radial distance loss to reduce the impact of branch size. By assigning larger weights to the centerline, CNNs pay more attention to the tiny structures. Zheng et al. [7] also adopt a distance-based weight in their loss function to alleviate the intra-class imbalance. However, these distance-based loss functions also give large weights to the voxels close to the skeleton, reducing the influence of breakages on the loss. To further resolve this problem, we design a breakage-sensitive loss that is not affected by the diameters of different branches.

For the second and third problems, Qin et al. [8] regard airway segmentation as a 26-channel connectivity prediction problem. But for the peripheral bronchi, the undetected voxels may account for half of the branch, which needs a larger receptive field and long-term features to learn the concept of connectivity. In contrast, for the conventional morphological methods [9, 10, 11], since they are merely designed to detect the tubular structure with a bright wall and a dark lumen, the performance can be more robust for the branches with deformed walls. To integrate this advantage into the CNN-based method, we adopt the geodesic distance transform and build a multi-stage framework in which one path highly relies on the intensity changes.

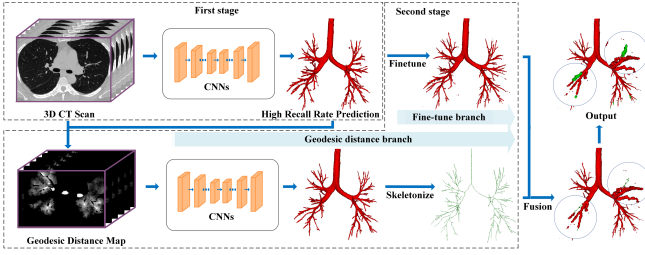

**Fig. 2**. Flowchart of our framework.

The main contributions of this paper are as follows: (1) A multi-stage segmentation framework is built with geodesic distance transform to alleviate the influence of the shape and intensity changes of the airway wall. (2) A breakage-sensitive loss is designed to reduce the number of breakages caused by intra-class imbalance. (3) Our segmentation results achieve state-of-the-art performance on two relatively large datasets, demonstrating the efficacy of the proposed method.

## 2. METHOD

An overview of the proposed framework is illustrated in Fig. 2. A two-stage training strategy is used. In the first stage, a network is trained to obtain high recall rate predictions. On this basis, the geodesic distance maps can be calculated. In the second stage, there are two branches. The pre-trained model is fine-tuned to improve accuracy in the fine-tune branch, while the geodesic distance maps are fed to another network for training in the geodesic distance branch. Finally, a fusion strategy is utilized to combine two outputs to get the final prediction result. More details about the multi-stage segmentation framework and breakage-sensitive loss are introduced in the following sections.

### 2.1. Multi-branch Segmentation Framework

Affected by blood vessels and lesions, the HU value of the bronchus wall and its surrounding area changes erratically. The shape and thickness of the airway wall also alter irregularly. Due to the small number of these samples, it is challenging for CNN-based methods to predict results with good continuity. However, no matter how the bronchus wall and its outer area change, the HU value in the cavity is always at a very low level. Therefore, conventional morphological methods are more suitable to solve this problem. Inspired by this, we adopt the geodesic distance transform (GDT) and propose a multi-stage framework (MSF) for airway segmentation.

We implement the geodesic distance transform by calculating the nearest geodesic distance from each voxel to the centerline voxel. To get accurate geodesic distance maps, we first train a network with a high recall rate in the first stage. Given a high recall rate result of the first stage network prediction $P$ from a CT scan, the largest connected domain and a centerline $C$ are extracted. By treating the voxels in the CT scan as the vertices of the graph and connecting them with the voxels in the 26 neighbors to form 26 edges, the CT scan is transformed into a 3D graph structure. The weight of the edge is defined as the sum of the gray values of the two vertices after mapping. Then we use a fast marching algorithm from scikit-fmm [12] to calculate the shortest path from each vertex on the centerline $C$ to all the other vertices on the CT scan and obtain the shortest path map $S$. Furthermore, the geodesic distance map $G$ is computed by

$$g_i = \min \{s_{i,j}, 1 \leq j \leq N\}. \tag{1}$$

where $g_i$ denotes the geodesic distance of the $i$-th voxel, $s_{i,j}$ denotes the shortest distance between the $i$-th voxel and the $j$-th centerline voxel, and $N$ is the number of centerline voxels. To reduce the influence of remote background voxels, the improved geodesic distance map $\hat{G}$ is further computed by

$$\hat{g}_i = \begin{cases} 0, & g_i \geq th, \\ th - g_i, & g_i < th. \end{cases} \tag{2}$$

Here, $th$ is a threshold used for truncation. The subtraction operation is to highlight the bronchi.

After getting the outputs of the two branches, an embedding fusion strategy is proposed to further reduce the breakages. Let $P_f$ denote the output of the fine-tune branch with higher precision and $P_g$ denote the output of the geodesic distance branch with better continuity. A threshold of 0.5 is firstly used to obtain the binary results. Then an operation to find the largest connected domain is performed on $P_g$ to remove the influence of outliers. By extracting the centerline map $C_g$ of $P_g$, the final result $P_r$ can be computed by

$$p_{ri} = \begin{cases} p_{fi} + p_{gi}, & if\ C^{-1}(p_{gi}) \in V, \\ p_{fi}, & else. \end{cases} \tag{3}$$

where $C^{-1}(\cdot)$ denotes the centerline voxel closest to the current voxel. $V$ is the set of all centerline voxels that have not been detected by $P_f$ and defined as

$$V = \{c_{gi} \mid p_{fi} = 0, c_{gi} \in C_g\}. \tag{4}$$

There are few breakages in the prediction of the geodesic distance branch. If the prediction result of the fine-tune branch does not include all the centerline voxels produced by the geodesic distance branch, it indicates that there may be fractures in its result. By embedding these missing centerline voxels and their adjacent voxels into the result of the fine-tune branch, the continuity of output will be improved.

**Table 1**. Comparison in the Binary Airway Segmentation Dataset(%).

| Method | Length | Branch | Precision |
|---|---|---|---|
| Jin et al. [1] | 85.38(10.39) | 83.07(11.49) | 93.93(1.92) |
| Juarez et al. [2] | 84.13(8.59) | 82.11(12.35) | 91.41(2.53) |
| Wang et al. [6] | 86.25(8.54) | 83.52(11.23) | 93.36(2.12) |
| Xue et al. [13] | 88.19(6.93) | 87.65(8.14) | 92.06(2.37) |
| Qin et al. [8] | 83.59(10.37) | 81.38(13.84) | **95.76(1.84)** |
| Qin et al. [3] | 91.82(5.31) | 87.64(9.18) | 91.54(2.89) |
| Zheng et al. [7] | 92.53(4.45) | 88.68(7.87) | 91.41(3.29) |
| Proposed | **95.33(2.78)** | **94.22(3.78)** | 91.84(2.80) |

**Table 2**. Comparison in our own Dataset(%).

| Method | Length | Branch | Precision |
|---|---|---|---|
| Jin et al. [1] | 81.71(12.31) | 80.52(10.68) | 94.69(1.87) |
| Juarez et al. [2] | 82.64(11.54) | 79.25(11.07) | 92.95(2.31) |
| Wang et al. [6] | 84.27(10.73) | 82.31(10.12) | 94.63(1.98) |
| Xue et al. [13] | 85.44(10.62) | 84.12(9.27) | 93.96(2.03) |
| Qin et al. [8] | 80.04(12.26) | 79.27(11.13) | **96.16(1.80)** |
| Qin et al. [3] | 89.03(6.89) | 85.43(9.18) | 92.87(2.61) |
| Zheng et al. [7] | 89.33(6.70) | 86.26(8.54) | 93.27(2.49) |
| Proposed | **92.21(4.03)** | **90.06(5.52)** | 93.78(2.16) |

### 2.2. Breakage-Sensitive Loss

Compared with the large bronchi, the volume of the distal airways is much smaller, which causes serious intra-imbalance. During training, the total loss may only drop a little when the predictions of the distal small airways become discontinuous. As a result, the breakages exist even in the prediction of the training set. To pose a high penalty for the breakages of the airways, we propose a breakage-sensitive loss (BS loss), which ignores the impact of branch size. The BS loss is defined as

$$L_{BS} = 1 - \frac{\sum_i^M p_i c_i}{\sum_i^M c_i + \varepsilon}. \quad (5)$$

where $p_i \in P$ denotes the prediction result of the i-th voxel, $c_i \in C$ denotes the i-th voxel value on the centerline map extracted from the label. $M$ is the number of the voxels of the CT scan and $\varepsilon$ is used to avoid dividing by zero. BS loss takes the centerline as the topological representation of the bronchus because the centerline is more representative to express the axial integrity of tubular structures. By neglecting the influence of bronchi diameters, BS loss can feedback a large gradient to the network to guide the network to predict results with fewer breakages when a fracture occurs.

General Union loss(GU loss) [7] is performed to supervise the segmentation results of each voxel and WingsNet [7] is used as the segmentation network. In the fine-tune branch, the total loss function can be written as

$$L_{total} = L_{GU} + w_t L_{BS}. \quad (6)$$

where $w_t$ is a hyperparameter and controls how much CNNs focus on the breakages. In the geodesic distance branch and the first stage, only GU loss is used for optimization.

## 3. EXPERIMENTS AND RESULTS

### 3.1. Dataset and Implementation Details

We evaluate our method in two datasets. The Binary Airway Segmentation Dataset [14] contains 90 CT images, 70 are from the LIDC [8], and the rest are from the training set of EXACT'09 [15]. The pixel spacing ranges from 0.5 to 0.821 mm and the slice thickness varies from 0.5 to 1 mm. We also collected 50 thin-slice CT images which are annotated by five experienced radiologists. The pixel spacing ranges from 0.5 to 0.873 mm and the slice thickness varies from 0.5 to 0.55 mm. The two datasets are randomly divided into the training set, validation set, and testing set at a ratio of 5: 2: 2 respectively. The model with the best validation results is used to test.

We truncate the HU value into the range between [-1000,600] and then mapped it to [0,255]. We sample 16 patches from each CT image with the size of [128,128,128]. Random Rotation around the x, y, or z axis is adopted with a probability of 0.5 in $[-15°, 15°]$. The skeleton is extracted using the algorithm in [16]. We train our model for 100 epochs. The learning rate is initialized to 0.01 and decays after the 60th and 90th rounds with a ratio of 0.1. In the second stage, $w_t$ is 0.5. Both the geodesic distance branch and fine-tune branch are trained for 30 epochs with a learning rate decay at the 15th and 25th epoch. The stochastic gradient descent (SGD) optimizer [17] with 0.9 momentum and $10^{-4}$ weight decay is used. All the networks are implemented in PyTorch 1.4 with an NVIDIA Tesla V100.

### 3.2. Evaluations

We evaluate our method with three evaluation metrics: precision, tree length detected rate [18], and branch detected rate [18]. Table 1 shows the results compared with other CNN-based approaches in the Binary Airway Segmentation Dataset. Jin et al. [1] use a a graph-based method to refine airways. Juarez et al. [2] combine dice loss and weighted cross-entropy loss to improve the sensitivity. Wang et al. [6] develop a radial distance loss to reduce the interference of branch sizes. Xue et al. [13] combine the prediction of the SDM with segmentation map in the training process to get better smoothness. Qin et al. [8] regard airway segmentation as a 26-channel connectivity prediction. They also adopt a feature recalibration module and an attention distillation module [3] to enhance the representation learning of bronchi. Compared with the above methods, our approach improves the rate of tree length detected and branch detected more than 2% and 5%, which reflects that our method outperforms the others in the detection of airways, especially small bronchi.

Table 2 also demonstrates the ability of our method to alleviate the fracture and detect more tiny branches. Its ex-cellent sensitivity also leads to a slight decrease in precision. This may be ascribed to: 1) Our model successfully detects some real thin airways, which are too fuzzy to be correctly annotated by experts. When calculating the evaluation metrics, these true airways are considered false positive, resulting in lower precision. 2) The bronchi wall of some end airway fades and can not be distinguished from the background texture. This situation brings a little leakage due to the BS loss's attention to small airways. However, it is clinically worthwhile to detect more tiny branches at such a trivial loss. Fig. 3 displays some results of our methods in 3D form. The approach in [7] is chosen as the baseline. Our framework improves the performance of the segmentation of the distal small airways and reduces the fractures. In the first example shown, it can be seen that the BS loss cannot completely solve the breakage problem compared with the baseline. But the result of the GDT has better continuity. The final output is improved based on the embedding fusion strategy.

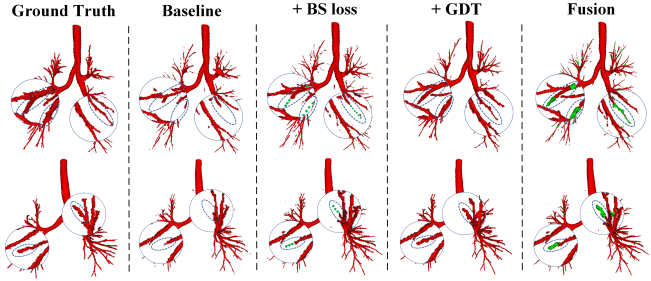

**Fig. 3**. The segmentation results of the proposed methods and the baseline. The green dots represent the skeleton points of the predictions of the GDT branch which are not included by the results of the fine-tune branch. The green patches are those voxels around the aforementioned skeleton points. Some breakages that cannot be resolved by BS loss are well patched after fusion.

**Table 3**. Ablation Experiment(%).

| Method | Length | Branch | Precision |
|---|---|---|---|
| Baseline | 92.53(4.45) | 88.68(7.87) | 91.41(3.29) |
| Baseline+BSL | 94.50(3.77) | 92.78(4.02) | **92.07(2.75)** |
| Baseline+MSF | 94.11(3.51) | 91.56(5.25) | 90.96(3.41) |
| Baseline+BSL+MSF | **95.33(2.78)** | **94.22(3.78)** | 91.84(2.80) |
| $w_t = 0.2$ | 92.71(4.98) | 90.35(5.82) | **93.05(2.61)** |
| $w_t = 0.5$ | 94.50(3.77) | 92.78(4.02) | 92.07(2.75) |
| $w_t = 1.0$ | **95.17(3.73)** | **92.98(3.94)** | 90.85(2.95) |
| Euclidean | 92.20(4.54) | 88.59(6.49) | 79.10(7.41) |
| Geodesic(gradient) | 90.99(10.52) | 88.57(12.00) | 91.56(2.96) |
| Geodesic(grayvalue) | **92.73(4.37)** | **89.72(6.93)** | **91.80(3.07)** |
| Simple addition | **95.99(2.70)** | **94.56(3.73)** | 89.61(3.34) |
| F2G | 94.38(3.09) | 92.33(4.61) | 91.14(3.11) |
| G2F | 95.33(2.78) | 94.22(3.78) | **91.84(2.80)** |

### 3.3. Ablation Study

We perform several ablation experiments in the Binary Airway Segmentation Dataset to further study the effect of each component in our segmentation framework. Quantitative results are demonstrated in Tabel 3. BS loss (BSL) can effectively boost sensitivity and precision by guiding the network to focus on breakage problems. Benefiting from the geodesic distance transform which can effectively utilize the intensity information of the image, the multi-stage segmentation framework (MSF) also enhances the sensitivity. Combining the two components, the results are further improved.

We also do ablation experiments on the fine-tune branch, the geodesic distance branch, and the fusion stage respectively. When $w_t$ changes from 0.2 to 1.0, the sensitivity increases while the precision decreases. This is because BSL can make CNNs pay more attention to fracture conditions. We further compare the performance when replacing the Geodesic distance with Euclidean distance. Due to the information of the image itself is not fully utilized, it is difficult to distinguish the lumen and wall of bronchi. Furthermore, two different methods of calculating geodesic distance are compared. When using the gradient as a measure of the geodesic distance between two voxels, the network fails in one case, resulting in a much higher standard deviation(10.52% and 12.00%). The reason is that the gradient is sensitive to noise and can be disturbed by the lesion easily. In contrast, employing the gray value of voxels to calculate the geodesic distance is more robust. Finally, three fusion strategies are evaluated. Simply adding the two predictions leads to a higher tree length detected rate and branch detected rate. But the precision drops a lot because this operation will make the airway thicker and produce more false-positive voxels. Using the result predicted by the fine-tune branch to repair that predicted by the geodesic distance branch (F2G), the fusion output gets worse. The reason is that the geodesic distance transform has better continuity in space and can produce results with fewer fractures than the finetune branch. On the contrary, adopting the result of the geodesic distance branch to repair the result of the finetune branch (G2F) is more appropriate.

### 4. CONCLUSION

In this paper, we introduce a multi-branch framework based on the geodesic distance transform to capture the intensity changes between airway lumen and wall and a breakage-sensitive regularization term to alleviate intra-class imbalance. With the geodesic distance transform and breakage-sensitive loss, CNNs can learn more continuity of the distal bronchi. Embedding fusion strategy is further used to reduce the breakages of airways. Extensive experiments show that our framework can extract more branches than other state-of-the-art methods while maintaining competitive segmentation performance.

## 5. ACKNOWLEDGMENTS

This research is partly supported by National Key R&D Program of China (No.2019YFB1311503),Committee of Science and Technology, Shanghai, China (No.19510711200).

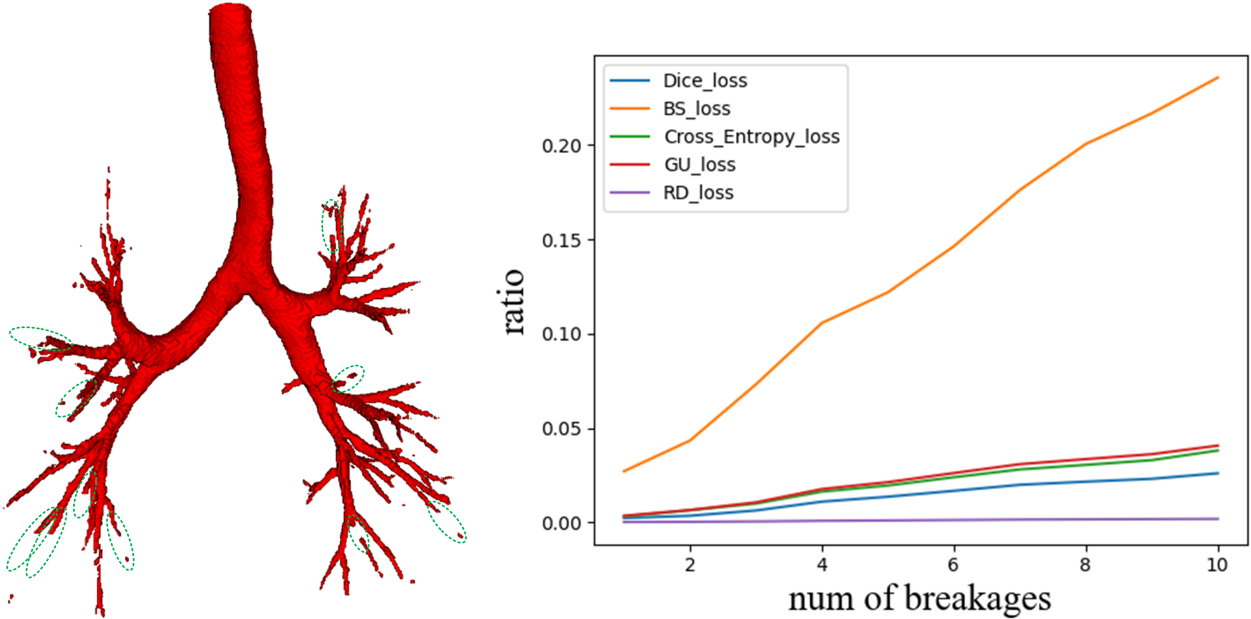

**Fig. 1**. The rates of change of different loss functions relative to their initial values when breakage occurs. All of the 10 fractures are circled by green dashed circles on the left. In the line chart on the right, the x-axis is the number of breakages of the distal bronchi and the y-axis is the rate of change of the loss function.

### A. MORE METHODOLOGICAL EXPLANATIONS

#### A.1. Difference with the Signed Distance Map

Signed Distance Map (SDM) is increasingly used in medical image segmentation tasks. Some researchers use distance map prediction as a regularizer during training to promote networks to learn more robust features for organ segmentation. Another researchers combine the prediction of SDM and the segmentation map to maintain the smoothness and continuity in shape. SDM calculates the distance from each voxel to the nearest boundary of the target organ and the sign denotes whether the voxel is inside the target organ (negative) or outside (positive). However, geodesic distance transform (GDT) counts on the weighted distance between each voxel and the closest centerline voxel. Instead of assigning a sign to represent the fuzzy relative position to boundary merely, GDT utilizes the intensity information of the image itself.

#### A.2. Sensitivity of different loss functions to breakages

To demonstrate BS loss's effectiveness, we compare it with other different loss functions in the ability to discover breakages, including Dice loss, Cross-Entropy loss, General Union loss (GU loss), and Radial Distance loss (RD loss). Let $l_i$ denote the value of the loss function when the i-th fracture occurs. Then the rate of change of the loss function $r_i$ can be obtained by

$$r_i = \frac{l_i - l_0}{l_0} \quad (1)$$

Fig. 1 illustrates the change of $r_i$ as the number of fractures increases. As we can see, BS loss is far more sensitive to fracture than other loss functions.

### B. ADDITIONAL QUALITATIVE RESULTS

Fig. 2 shows that our proposed method has fewer fractures in the form of 2D slices. Some cases of false positives are shown in Fig. 3. After a retrospective evaluation of the label, some false positives are identified as true airways. Due to the difficulty of annotation, these branches are inadvertently ignored.

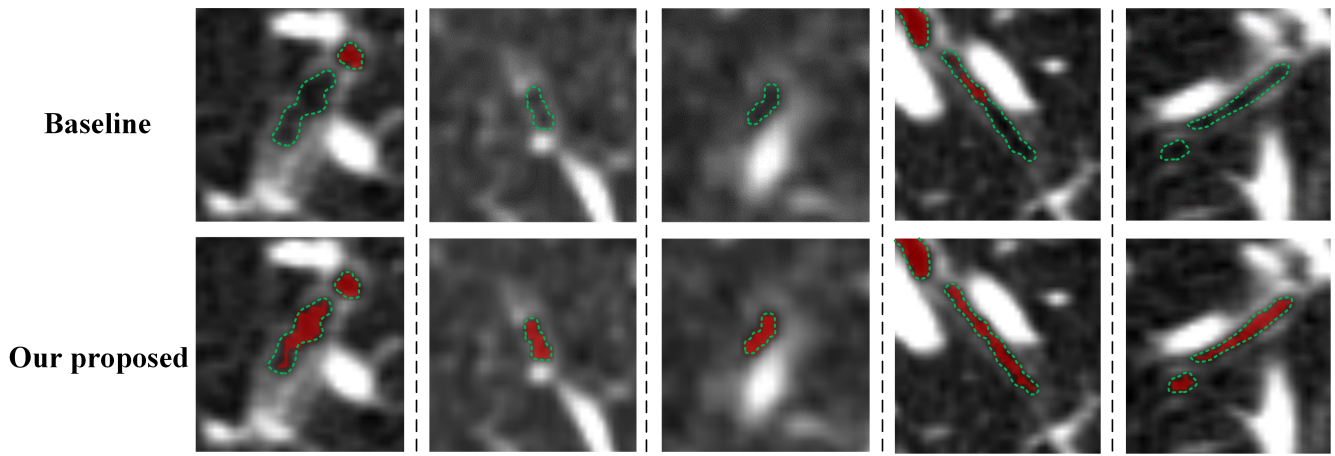

**Fig. 2**. The segmentation results of the proposed methods and the baseline in 2D slices. The prediction of the network is in red and the label is indicated by the green dotted line.

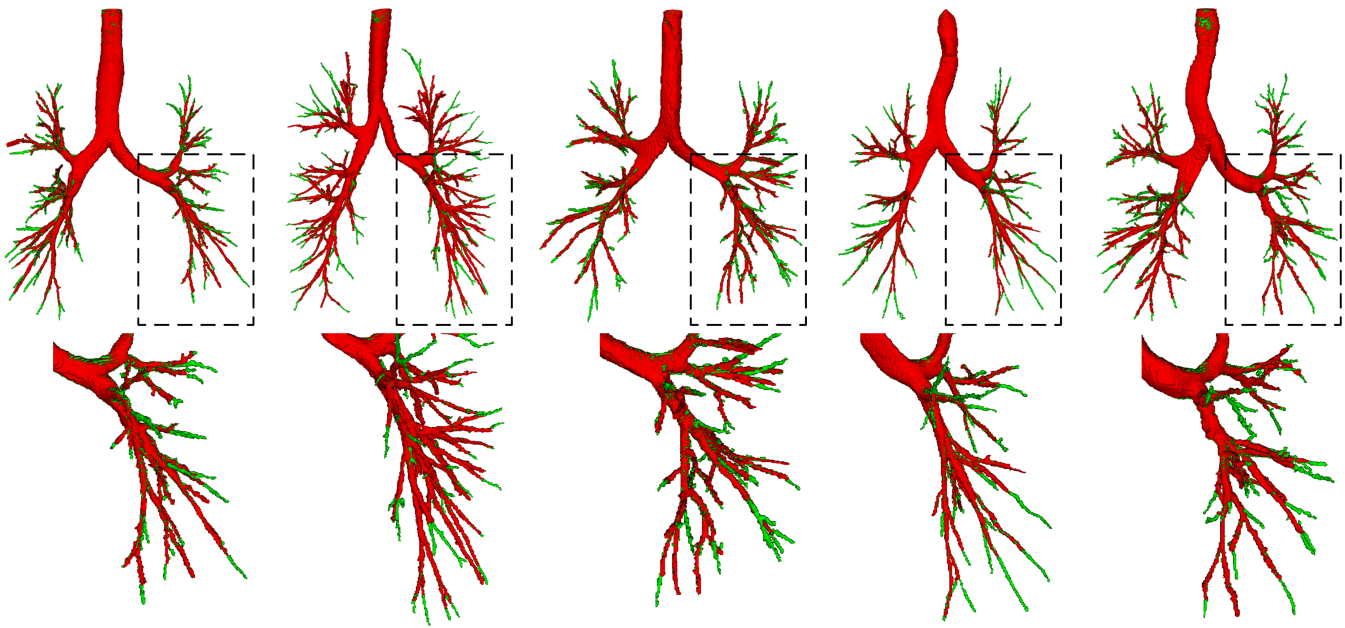

**Fig. 3**. Examples of false positive. The true positive voxels are red and the false positive voxels are green. Some false positives are identified as true airways after retrospective evaluation of labels. These branches are unintentionally neglected due to annotation difficulty.